# Cell surface deformation during an action potential

C. Fillafer, M. Mussel, J. Muchowski, M.F. Schneider
Medical and Biological Physics
Faculty of Physics
Technical University Dortmund

matthias-f.schneider@tu-dortmund.de

## Abstract

The excitation of many cells and tissues is associated with cell mechanical changes. The evidence presented herein corroborates that single cells deform during an action potential (AP). It is demonstrated that excitation of plant cells (*Chara braunii* internodes) is accompanied by out-of-plane displacements of the cell surface in the micrometer range (~1–10 μm). The onset of cellular deformation coincides with the depolarization phase of the AP. The mechanical pulse *(i)* propagates with the same velocity as the electrical pulse (within experimental accuracy; ~10 mm s$^{-1}$), *(ii)* is reversible, *(iii)* in most cases of biphasic nature (109 out of 152 experiments) and *(iv)* presumably independent of actin-myosin-motility. The existence of transient mechanical changes in the cell cortex is confirmed by micropipette aspiration experiments. A theoretical analysis demonstrates that this observation can be explained by a reversible change in the mechanical properties of the cell surface (transmembrane pressure, surface tension and bending rigidity). Taken together, these findings contribute to the ongoing debate about the physical nature of cellular excitability.

## Introduction

Action potentials (APs) are intriguing phenomena that appear in many biological systems (neurons, myocytes, excitable plant cells, etc.). For a long time, it has been believed that these pulses are of an electrical nature. The mathematical description of APs was based on the view that the excitable membrane can be fully represented by an equivalent circuit (1). However, this approach has come under criticism. The debate has been stirred up by Tasaki (2–10) and has been extended mainly through the works of Kaufmann (11, 12) and Heimburg (13, 14). One of the central points of criticism of the electrical framework is that it neither contains nor predicts non-electrical manifestations of the AP. These pulse components, however, exist and include optical (2), thermal (15), magnetic (16) as well as mechanical (3–6, 17) changes at the cell surface. The latter have been studied with a variety of highly sensitive techniques (piezoelectric benders, interferometry, AFM). The inherently soft nature of nervous tissue preparations combined with the sheer minuteness of the movements posed difficulties and led to varying results (3–5, 17, 18). Nevertheless, more recent studies are widely in agreement and indicate that the mechanical pulse in cylindrical axons is of biphasic nature with expansion followed by contraction (~1–10 nm) (6, 19). In parallel, there exists a biphasic intracellular pressure wave (19). Intriguingly, neither the mechanism behind the mechanical pulse component nor its relation to the electrical events are currently understood. Aside from APs in axons, other excitation phenomena in biology are also associated with mechanical changes. Deformations were reported most prominently in muscle cells (20) but also during other excitation phenomena such as spreading depression waves in cortical tissue (21). It must be of central interest to investigate if these phenomena can be explained by a unified theory.

While there exists firm evidence that an AP is *not only* an electrical *but also* a mechanical pulse, several open questions remain. Even in well-cleaned axons the cell surface is covered by extracellular matrix and Schwann cells (7). This makes it difficult to observe and study the excitable membrane directly. Moreover, if this sheath is stiffer than the underlying cell membrane it will lead to significant attenuation of mechanical signals. Thus, the actual mechanical changes during an AP may be larger than anticipated. Herein, we attempt to contribute to these open problems. Mechanical changes are investigated during AP propagation in plant cells. Internodes from *Charophytes* are well suited for this purpose. These cells are large (diameter ∼0.5—1 mm; length ∼1—15 cm), easy to handle and have a long-standing history in excitable cell research (22, 23). In *Charophytes*, comparatively large radial and axial deformations have been reported during an AP (~100 nm (4, 24)). As in the case of axons, the mechanical pulse is biphasic and consists of expansion followed by contraction (25). Herein, we demonstrate that "freeing" the excitable membrane from the constraints of the plant cell wall reveals even larger surface displacements in the micrometer range. A theoretical analysis indicates that these deformations during an AP can be attributed to reversible changes in three mechanical properties of the cell surface (transmembrane pressure, surface tension, bending rigidity).

## Materials and Methods

*Materials.* All reagents were purchased from Sigma-Aldrich (St. Louis, MO, USA) and were of analytical purity (≥99%). Glass capillaries were obtained from Sutter Instrument (Novato, CA, USA).

***Cell cultivation and storage.*** *Chara braunii* cells were cultivated in glass aquariums filled with a layer of 2-3 cm of New England forest soil, quartz sand and deionized water. The cells were grown under illumination from an aquarium light (14W, Flora Sun Max Plant Growth, Zoo Med Laboratories Inc., San Luis Obispo, CA, USA) at a 14:10 light:dark cycle at room temperature (~20°C). Prior to use, single internodal cells were stored for a minimum of 12 h in a solution containing 0.1 mM NaCl, 0.1 mM KCl and 0.1 mM $CaCl_2$.

***Plasmolysis of Chara internode.*** A single internodal cell (3-6 cm long) was placed on a Plexiglass frame into which compartments (~2 x 5 x 10 mm; h x w x l) had been milled. The bottom of the frame consisted of a glass coverslip. Small extracellular sections (length ~5 mm) of the cell were electrically isolated against each other with vacuum grease (Dow Corning Corporation, Midland, MI, USA). The grease also provided structural support for the cell during plasmolysis. Artificial pond water was added (APW; 1 mM KCl, 1 mM $CaCl_2$, 5 mM HEPES, 110 mM D-sorbitol; pH set to 7.0 with NaOH). This APW was replaced gradually with APW of higher osmolarity (regulated by addition of D-sorbitol; initial: ~120 mOsm $kg^{-1}$; final: ~270 mOsm $kg^{-1}$). Addition of ~0.5% bovine serum albumin to the final APW was crucial to minimize adhesion between aspiration pipette and cell membrane. After an equilibration time of 30-60 min, the plexiglass chamber was fixed on the microscope stage. A waveform generator (Agilent

33250A; Agilent, Santa Clara, CA, USA) in combination with a stimulus isolation unit (SIU5; Grass Technologies, Warwick, RI, USA) was used to trigger APs. The membrane potential in one of the compartments far from (1-5 cm) the stimulation site was monitored by intracellular recording. Deflections of the edge of the cell surface are presented as kymographs. In brief, the intensity profile along a line was extracted from every frame of the video recordings (framerate: 10-40 s$^{-1}$) and was assembled in ImageJ (Version: 1.46r; http://imagej.nih.gov/ij) using the macro ImageJ Kymograph (by J. Rietdorf and A. Seitz). Brightness and contrast of the final kymographs was adjusted. Prominent features in the intensity profile correspond to cell membrane and protoplast edge respectively. The latter is particularly contrast rich due to the presence of chloroplasts. The membrane potential recording was temporally synchronized with video microscopy by an LED flash into the optical path of the inverted microscope (Olympus IX71). The time difference between the membrane potential pulse and the mechanical displacement was calculated by depicting deviations in both signals. The criterion for the time of arrival at the measurement site was defined as a deviation of the signal from baseline by three times the standard deviation of baseline variance.

*Micropipette aspiration during an AP.* A hypodermic needle was used to make a small incision in the cell wall cylinder of a plasmolysed *Chara* internode (26). The medium in the first compartment was replaced with 150 mM KCl to facilitate membrane potential recording via the K$^{+}$-anesthesia technique. Glass pipettes were pulled to a needle tip (P-97 micropipette puller; Sutter Instrument, Novato, CA, USA) and were broken after scoring with a second pipette to obtain a flat tip. The pipette was filled with APW and was connected to a water column whose height was regulated by a micromanipulator. For measurements of the membrane potential, a Ag/AgCl electrode was looped through the water column into the aspiration pipette (PS-2132; 50Hz sample rate; PASCO scientific, Roseville, CA, USA). The typical technical requirements and procedures for micropipette aspiration can be found in the literature (27, 28). When aspirating the membrane of plasmolysed protoplasts, one deals with a less defined cellular geometry. Oftentimes, the convoluted protoplast shapes made it difficult to observe the region in which the aspiration pipette touched the cell membrane (because a lobe of protoplast obscured the initial 10-50 µm of the pipette). In such cases, it was possible to reveal the membrane projection by aspirating it beyond the overlapping region. However, this procedure prolonged the experiment and in general required higher suction pressures. To ensure comparability of the results, we resorted to the following procedure: Care was taken to find a position of the protoplast edge, at which the point of contact between pipette and protoplast surface was directly observable. The pipette was slightly pressed against the membrane and a suction pressure was applied. This pressure was insufficient $0 < \Delta p < \Delta p_{cap}$ to aspirate a membrane projection that is longer than the pipette radius ($i.e., L_p > R_p$). The negative pressure required to meet this condition was in the range of ~$10^3$-$10^4$ mN m$^{-2}$ ($R_p \sim 10\ \mu m$). The system was allowed to equilibrate for several 10 sec. Subsequently, two spatially distant electrodes

(several mm to cm away) were used to trigger an action potential that propagated past the aspiration site.

***Mechanical model of micropipette aspiration.*** Complex aspiration scenarios have been studied previously (29, 30). However, these works were carried out in a different context, in a different regime of the parameter space and did not directly focus on the question posed herein. During an AP, the aspirated *Chara* cell does not reach a new stable mechanical state. Thus, our focus is only on identifying the stability conditions for the weakly aspired state (zero aspiration). Such compromise allows the use of a simplified spherical geometry. The model assumptions are: (*i*) only the simplest surface contributions are considered: surface tension ($\sigma$) and the linear regime of the bending rigidity ($\kappa$) (31). While we have focused here on the simplest elastic description of a cell surface, additional parameters may also change during an AP (*e.g.*, spontaneous curvature, in-plane shear rigidity and surface compressibility). It may be worth to investigate a more elaborate model of cell surface elasticity in future work. (*ii*) At equilibrium, there are no internal flows in the bulk or along the surface; *i.e.*, statics implies that $\sigma$, $\kappa$, and $\Delta p$ are constants. In principle, these parameters should be coupled to one another by a state equation. Their dependence on geometrical factors (*e.g.*, surface area) was neglected, because the focus was on the initiation of the instability and not on determining the final strongly-aspirated state. (*iii*) A simplified geometry of a spherical cell was considered instead of the cell-wall-bounded cylinder, because it allows a rather simple analytic expression of the energy function. The simplification is reasonable because the cell volume is much larger than the aspirated segment, $V_{cell} > 10^6 V_{asp}$. The volume of the sphere was matched to that of a plasmolysed *Chara* cell by setting $R = 10^2 R_p$, with $R_p = 10$ $\mu$m. The model geometry is depicted in Fig. 1, although not to scale (the pipette radius is a hundred times smaller than the cell radius). (*iv*) Changes in cell volume during an AP were assumed to be negligible since $\frac{\Delta V}{V} \sim 10^{-4}$ (32). (*v*) The pressure inside the pipette $p_p$ was assumed constant.

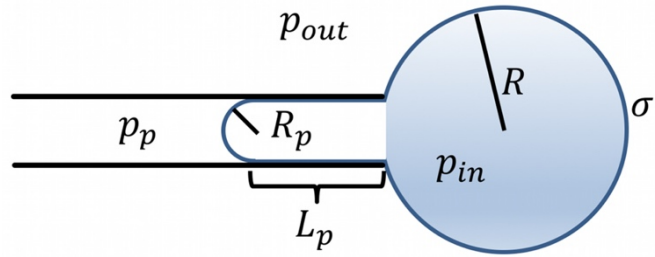

**Figure 1.** Geometry of the aspiration model (not to scale; in the calculations $R/R_p \sim 10^2$).

The favorable shape was calculated by minimizing the elastic energy function
$$E = \sigma \int dA + \frac{\kappa}{2} \int (2H)^2 dA + p \int dV \quad (1)$$
with *A*, the surface area, *H*, the mean curvature of the surface and *V*, the cell volume. For the simplified geometry considered, the energy function is
$$E = \sigma A + \kappa g_{curv} + p_{out} V_{sph} + p_p V_{asp} - p_{in} V_{tot}. \quad (2)$$
The area and curvature contributions to the energy are respectively

$$A = 2\pi R_p^2 + 2\pi R_p L_p + 2\pi R^2[1 + \cos(\alpha)], \quad (3)$$

$$g_{curv} = 4\pi \left\{ 4[1 + \cos(\alpha)] + \frac{L_p}{R_p} + 4 \right\}, \quad (4)$$

with

$$\sin(\alpha) \equiv \frac{R_p}{R}. \quad (5)$$

The volume contribution is partitioned into $V_{sph}$ the volume of the part of the cell outside of the pipette (shaded area in Fig. 1), $V_{asp}$ the volume enclosed in the pipette, and $V_{tot} = V_{sph} + V_{asp}$ (30).

$$V_{sph} = \frac{2\pi}{3} R^3 [1 + \cos(\alpha)] + \frac{\pi}{3} R_p^2 R \cos(\alpha)$$

$$V_{asp} = \frac{2}{3} \pi R_p^3 + \pi R_p^2 \quad (6)$$

The assumption of a constant cell volume simplifies the energy expression into

$$E = \sigma A + \kappa g_{curv} + \Delta p V_{sph} + const, \quad (7)$$

with $\Delta p = p_{out} - p_p$.

## Results

***Cell surface deformation during an AP.*** In a native *Chara* cell, the plasma membrane is tightly pressed against the cell wall by a turgor pressure (~6·10⁵ N m⁻² (32)) (Fig. 2a). However, by increasing the extracellular osmolarity it was possible to progressively reduce turgor until the plasma membrane detached from the cellulose sheath – a process known as plasmolysis (Fig. 2b) (26, 33). During this procedure, the protoplast did not retract uniformly. In certain regions (*e.g.,* at the nodes) the membrane still adhered to the cell wall whereas in other areas it detached entirely. Initially, the shape of a plasmolysed cell was irregularly wavy. As time progressed, the protoplast equilibrated, assumed an unduloid-like form, and eventually fragmented. This process was described previously (28) and resembles a pearling instability that develops at a very slow pace. In the medical/biological literature this phenomenon is sometimes referred to as *beading* or *varicose*. Physically, it is related to the Plateau-Rayleigh instability.

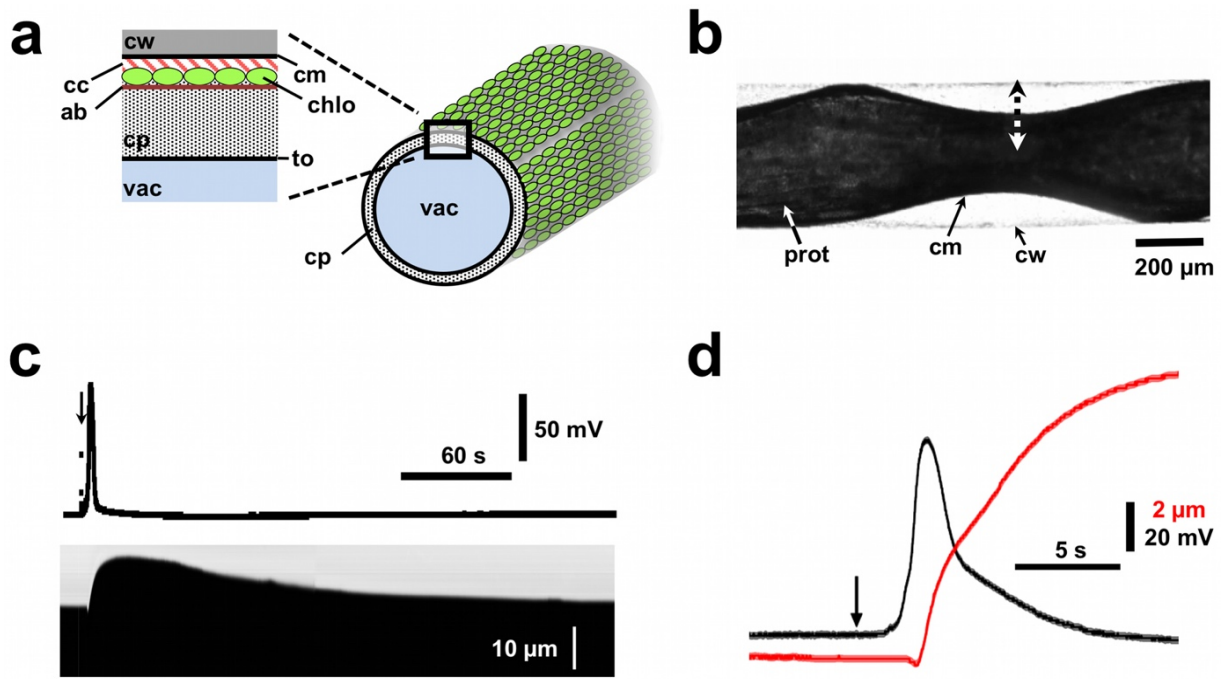

**Figure 2. Cell surface deflection during an action potential (AP). (a)** In *Chara*, the cytoplasm (cp) is marginalized by the tonoplast (*to*)-covered vacuole (*vac*). The cellular cortex consists of the cell wall (*cw*), cell membrane (*cm*), cortical cytoskeleton (*cc*), chloroplasts (*chlo*) and subcortical actin bundles (*ab*) (see (34)). **(b)** When turgor was reduced by increasing the extracellular osmolarity, *cm* separated from *cw*. Deflections (dashed arrow) of the projection edge of the protoplast surface (*prot*) were tracked by light microscopy. **(c)** Upon excitation of an AP the cell surface underwent a biphasic, reversible deflection (stimulus indicated by arrow; top trace: membrane potential; bottom trace: kymograph of surface deflection). **(d)** Membrane potential pulse (black) and out of plane displacement of the cell surface (red); note: an initial inward movement is followed by expansion.

*Chara* cells did not lose excitability in the course of plasmolysis. Thus, it was possible to stimulate APs and to study if deflections of the cell surface occur. For this purpose, randomly chosen regions of the protoplast edge were tracked by light microscopy (Fig. 2b). In the absence of electrical stimulation only minor drift of the edge was observed (Fig. S1 in the Supporting Material). In contrast, a distinct surface displacement occurred upon excitation of an AP (in 142 out of 152 cases; $N$=30 cells) (Fig. 2c and Video S1 and S2). The maximum deflection was more often outward (95 cases) than inward (47 cases) and typically in the 1—10 μm range (Fig. S1). In the majority of experiments a brief displacement (≤1 s) with opposite directionality preceded the maximum deformation (109 cases; Fig. 2d and Video S2). Such biphasic displacements were also reported in fully turgid cells, albeit with 10—100x lower amplitudes (25). In general, the amplitudes and time courses of the deformations were quite variable at different locations along the protoplast projection edge (Fig. S1). This variation will be discussed in a forthcoming manuscript.

***Correlation between membrane potential and surface displacement pulse.*** The AP propagation velocity calculated from the mechanical displacement (8.2±2.4 mm s$^{-1}$; $n$=26 pulses in $N$=4 cells), within experimental accuracy, agreed with that based on the electrical pulse (9.6±2.0 mm s$^{-1}$; Fig. S2). In most experiments the surface displacement slightly trailed the membrane potential pulse (Fig. 2d). However, since the electrical measurement is by default not as localized as a mechanical measurement, the delay between the electrical and mechanical pulse may be a measurement artifact. Future studies could circumvent this difficulty, for instance, by using fluorescent imaging which allows for localized monitoring of the membrane state. In any case, it was evident that the mechanical deformation outlasted the electrical component (Fig. 2c). The membrane potential pulse in a plasmolysed cell usually had a duration of ~10—20 s, whereas the surface deflection relaxed on timescales that were an order of magnitude longer (~0.5—5 min). The latter agrees with observations in fully turgid *Charophytes* (24, 25). It is of interest to note that the first derivative of the displacement with respect to time (*i.e.* the out of plane velocity of the membrane), correlates fairly well with the membrane potential pulse (Fig. S3). This is an indication that the two pulses are not independent phenomena.

***Involvement of $Ca^{2+}$ and actin-myosin-motility in surface displacement.*** Actin and myosin are present in *Characean* cells, where coherent sliding of myosin-coated organelles on actin filaments leads to directional streaming of the cytoplasm (velocities up to 100 μm s$^{-1}$) (35). During an AP, streaming is temporarily arrested (excitation-cessation-coupling) and recovers within several minutes. This resembles the relaxation time scales of the surface deformation as observed herein and by others (*cf.* Fig. 2 and Ref. (25)). Thus, the cellular events that trigger stoppage of streaming (probably an increase of intracellular $Ca^{2+}$ (36)) or a coincident process in the cytoskeleton may be involved in the observed deformations. We attempted to investigate these possibilities. It has been reported that membrane excitation can be uncoupled from the cessation of streaming by replacing extracellular $Ca^{2+}$ with $Mg^{2+}$ (37). This approach, however, was not feasible, because plasmolysis in the absence of $Ca^{2+}$ led to rupture of the *Chara* cell membrane. Another technique that was proposed to uncouple membrane excitation from streaming in *Nitella* – addition of manganese chloride (5 mM) to the external medium (38) – was also not successful as it led to spontaneous activity and deterioration of the cell. We finally attempted to gradually reduce the extracellular concentration of $Ca^{2+}$ by addition of a chelating agent (EGTA, 0.1-2 mM). In this concentration range, the stoppage of cytoplasmic streaming during an AP persisted. Upon addition of increasing amounts of EGTA (10 mM), the membrane potential record started to fluctuate, streaming became irregular and the cell exhibited signs of deterioration. Thus, complexation of extracellular $Ca^{2+}$ by addition of EGTA is not a viable procedure to decouple membrane excitation from stoppage of cytoplasmic streaming in *Chara*. The replacement of extracellular Ca2+ by other cations ($Mg^{2+}$, $Mn^{2+}$, $Ba^{2+}$) were also tested on *Nitella translucens,* but it was not possible to reach a state in which the cell was plasmolyzed and excitable.

In a subsequent series of experiments, cytochalasin D (CytD) was employed. Cytochalasins have been reported to uncouple membrane excitation and contraction in muscle cells (*i.e.* the membrane is excitable, but contractility is impaired (40)). These substances also interfere with and arrest cytoplasmic streaming by a presently unknown mechanism (39). Thus, we hypothesized that treatment with CytD will abolish the cell surface deformation in *Chara* if the latter is dependent on actin-myosin motility.

When a *Chara* cell was incubated with CytD, cytoplasmic streaming came to a halt but the cell remained excitable. This is analogous to the effects of CytD on muscle cells. When an AP was triggered under these conditions, the surface deformation of the *Chara* cell persisted (Fig. 3). The time scales of the deflection were similar and the amplitudes were slightly larger as compared to native cells. This is in contrast to the observations of diminished contractility with CytD in muscle cells (40). These results provide some indication that actin-myosin-motility as in muscle cells is not involved in the surface deformation during an AP in *Chara*.

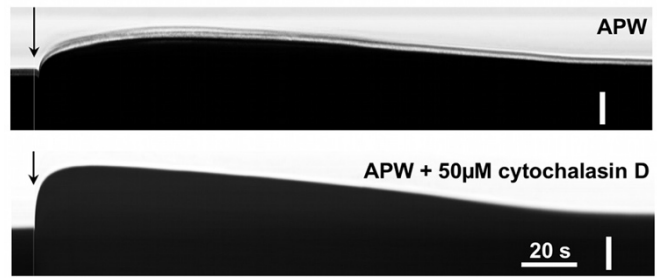

**Figure 3. Effect of cytochalasin D on surface deflection during an action potential (AP). (top)** Displacement of the cell surface upon excitation of an AP in artificial pond water (APW) and **(bottom)** in a different cell that had been incubated with APW + 50 μM cytochalasin D. Stimulus indicated by arrow. Vertical scale bars represent 20 μm.

*Micropipette aspiration at rest and during an AP.* To better understand the mechanism of the cell surface displacement, micromechanical tests were carried out. A small incision (length ∼500 μm) was made in the cell wall of a plasmolysed *Chara* internode. Through this opening it was possible to directly access the cell surface (see Fig. 2a and Ref. (34)).

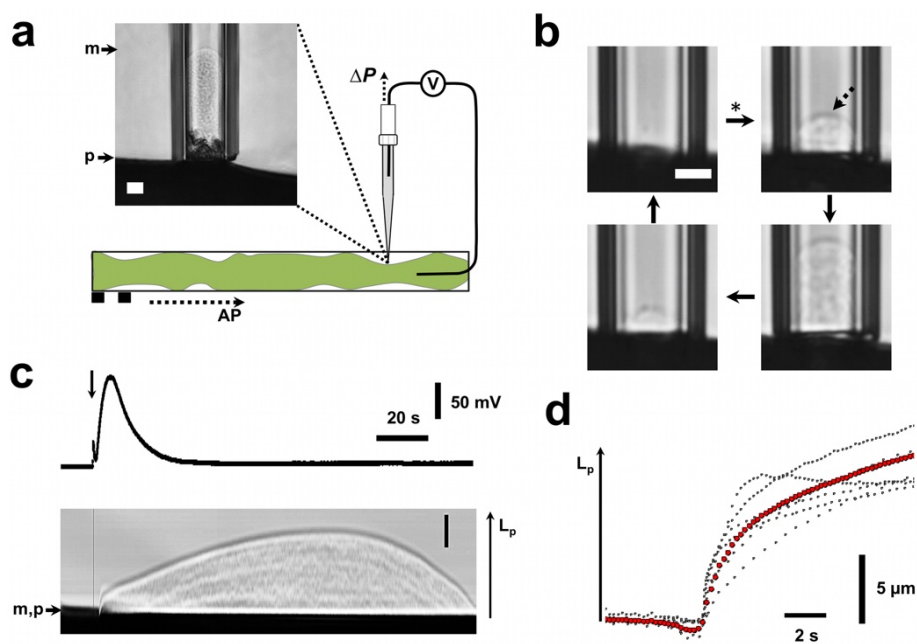

**Figure 4. Cell mechanical changes during an action potential (AP). (a)** Aspiration of *Chara* cell membrane into a micropipette (membrane projection indicated by ***m***, protoplast surface by ***p***). Note: the cell membrane is peeled off the dense array of chloroplasts (also see Video S3). **(b)** During an AP the membrane underwent a reversible cycle of motion into and out of the pipette at constant suction pressure. Aspirated membrane cap is indicated by a dashed arrow **(c)** Suction pressure (*Δp*) prior to stimulation of an AP was "clamped" at 0<*Δp*<*Δp*$_{cap}$ (see text for definition of *Δp*$_{cap}$). Membrane potential record (top) and aspirated length (*L*$_p$; bottom) during an AP. **(d)** Initial phase of membrane motion into pipette (*n*=6 experiments in *N*=4 cells; individual traces (black) and average (red)). See text for additional data and statistics. Unlabeled scale bars represent 10 μm.

Micropipette aspiration was used to study the mechanical properties of this surface. For a cylindrical cell that is aspirated into a pipette, the surface tension ($\sigma$) is given by the Young-Laplace law

$$\sigma = \Delta p \left( \frac{2}{R_p} - \frac{1}{R_c} \right)^{-1}$$

with *Δp* the pressure difference between extracellular medium ($p_{out}$) and pipette ($p_p$), and $R_c$ and $R_p$ as the radii of the cell cylinder and the pipette respectively (27). The cortical tension of the cell in the resting state $\sigma_{rest}$ (*i.e.*, before excitation) was determined from the pressure difference ($\Delta p_{cap}$) that is required to aspirate a membrane projection with length ($L_p$) equal to the pipette radius (*i.e.*, when a hemispherical membrane cap was aspirated (28)). In plasmolysed *Chara* cells, $\sigma_{rest}$ was 0.06±0.01 mN m$^{-1}$ (*n*=4 experiments in *N*=3 cells; *Δp* ~10 N m$^{-2}$, $R_c$~150 μm, $R_p$~10 μm). This value of membrane tension is in good agreement with that of other excitable systems, for instance, *Nitella* protoplasmic droplets (~0.05 mN m$^{-1}$ (41)) and molluscan neurons (~0.04 mN m$^{-1}$ (42)).

For aspiration experiments during an AP, the pipette touched the cell surface and a slight suction pressure was applied ($0 < \Delta p < \Delta p_{cap}$; see Fig. 4 and Methods for details). Once the position of the membrane projection within the pipette had remained relatively steady, an AP was induced several mm to cm away and propagated past the aspiration site. Although the pipette pressure was held constant, $L_p$ increased upon arrival of the AP in all experiments conducted (*n*=22; *N*=9 cells; Fig. 4, Video S3). In some cases, a short "inward dip" occurred prior to this movement (Fig. 4d). The membrane projection either moved into the pipette irreversibly (Fig. S4) or reached a maximum and relaxed back to its initial position (Fig. 4c and Video S3). Since cell surface deflections in absence of a pipette were reversible (Fig. 2), it seems likely that irreversibility was a concomitant of the aspiration procedure (29). In general, reversibility prevailed (14 out of 22 cases).

***Mechanical analysis of micropipette aspiration during an AP.*** In a typical aspiration experiment, the surface forces, which are applied by the surface tension $\sigma$ and the bending rigidity $\kappa$, balance the pressure difference between the extracellular medium and the interior of the pipette (*Δp*). This triplet of surface properties ($\Delta p, \sigma, \kappa$) represents the mechanical state of the system in a 3d-phase space. The objective of this section is to identify the conditions under which the balance of forces of a weakly aspirated cell is disrupted, such that the system is progressively aspirated (Fig. 4). This was achieved by calculating the aspiration length ($L_p$) that minimizes the elastic energy of the cell surface for different values of the mechanical parameters (for details see Eq. 7). An example of the energy as a function of $L_p$ is provided for three values of the surface tension with the other parameters ($\Delta p$ and $\kappa$) held constant (Fig. 5a). This graph demonstrates the existence of a critical value of $\sigma$, that flattens the energy function. From there, an increase of surface tension stabilizes the weakly aspirated state ($L_p = 0$), while a decrease of $\sigma$ leads to an instability; i.e., $L_p$ increases with time, which means that the cell is aspirated into the pipette.

The phase space of the system consists of two regimes: one in which the weakly-aspirated state is stable and one where it is not. A state is unstable when the surface tension and bending rigidity are insufficient to balance the pressure difference (28). A 2d-slice in the $\Delta p - \sigma$ plane of the phase-space is depicted in Fig. 5b. Weakly aspirated states are stable below the dashed line (instability line). The estimated resting state of a plasmolysed *Chara* cell is located in this regime and is marked by a grey ellipse. A transition across the line into the unstable regime (from stable $L_p = 0$ to $L_p > 0$) can be induced, for example, by increasing $\Delta p$ by ~5 N m⁻², decreasing $\sigma$ by ~50% or decreasing $\kappa$ by 2-3 orders. The effect of the latter is much smaller, and requires a close proximity of the initial state to the instability line. The short inward motion of a weakly aspirated cell at the beginning of an AP (Fig. 4d) can be induced by a parameter change with opposite directionality (i.e., a decrease in $\Delta p$ or an increase in $\sigma$ or $\kappa$). These findings are in line with a more elaborate calculation conducted for a different domain of the parameter space (29).

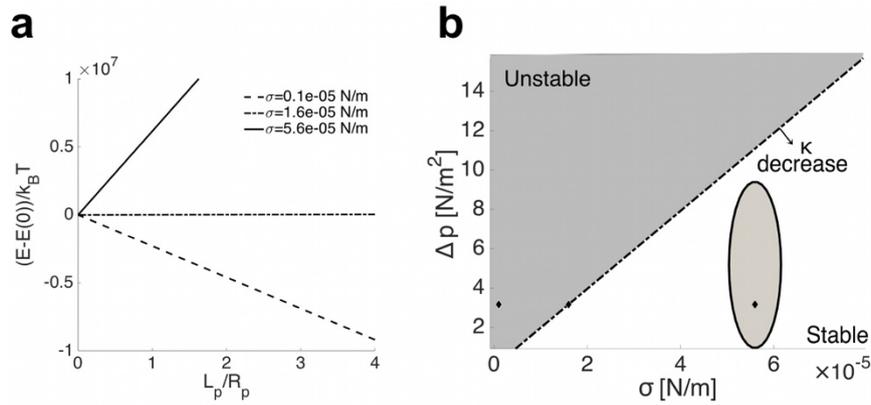

**Figure 5. (a)** The energy $E$ as a function of aspiration length $L_p$ for three values of the surface tension reveal that the weakly aspired state ($L_p = 0$) can be stable, critical and unstable. Other parameters were held constant: $\kappa = 10^{-19} J$, $\Delta p = 3 \frac{N}{m^2}$, $R_p = 10$ μm. For convenience, the energy was scaled to zero at an aspiration length of zero ($L_p = 0$). In addition, it was normalized by the mean thermal energy at room temperature ($20^o C$), to indicate that the elastic energy stored in the surface is considerably larger. **(b)** The instability line in the $\Delta p$–$\sigma$ phase space. Weakly aspirated states are located below the line. The estimated resting state of the cell is depicted by the grey ellipse (its width reflects the experimental error in the measurement of $\sigma_{rest}$, whereas its height reflects the experimental error in the aspiration pressure that sets the weakly aspirated state). Decreasing $\kappa$ effectively shifts the instability line in the direction of the arrow. The three states studied in (a), are depicted as small diamonds in (b).

## Discussion

It has been documented by several independent investigators that excitation of cells and tissues is accompanied by a deformation (2–10, 17, 19, 21, 24, 25). Mechanical pulses with micrometer amplitudes have been reported, for instance, upon excitation of myocytes (1) and during spreading depression (SD) waves in cortical tissue (21). During an AP in axons, the deformations are typically on the scale of 1—10 nm, whereas in fully turgid *Charophytes* they reach ~100 nm. Herein, it was demonstrated that sheathing material and/or cells attenuate these motions. When the plasma membrane of a *Chara* cell was separated from the rigid external cell wall, considerably larger displacements were observed (1-10 μm compared to ~100 nm; Fig. 2). The present work and the majority of studies of cell surface deformations (9, 19, 21, 25, 43) are in agreement that the biological material (cell, nervous tissue, etc.) deforms in a biphasic manner (swelling followed by contraction or *vice versa*). Up to date, it is not known if these deformations in myocytes, axons, excitable plant cells and cortical tissue can be explained by a unified theory.

***On the relation between the electrical and mechanical events.*** There are at least three potential relations between the membrane potential pulse and the cellular deformation: *(i) The mechanical and electrical components belong to the same phenomenon (i.e.* one observes two realizations of the action potential). An obvious objection to this viewpoint comes from considering the different behavior in time of membrane potential and the surface deformation. In *Chara*, the cell surface recovers its shape on time scales that are roughly an order of magnitude longer than the membrane potential pulse (30—300 sec versus 5—10 sec). However, a difference in time course does not in itself mean that one deals with two different phenomena. Such behavior of variables is common, for example, in coupled differential equations. In this regard, it is of interest to note that the mechanical relaxation time agrees fairly well with the duration of the relative refractory period (> 60 sec at room temperature (44)). Thus, it could be that the mechanical observable reflects state changes of the excitable medium in a more direct manner as compared to the electrical one. This view can be falsified if there is deformation disappears when there only is an AP at the cell membrane (*i.e.* when no other phenomena such as stoppage of cytoplasmic streaming and tonoplast AP take place). *(ii) The cell surface deformation is an independent phenomenon triggered by the AP.* This is equivalent to the current view of excitation-contraction coupling in muscle cells. There, an AP at the plasma membrane is believed to trigger cellular shortening via a process which involves intracellular polymers (1). This shortening lags behind and outlasts the electrical pulse (20). Similarly, an AP in *Chara* affects an intracellular process based on filaments (cytoplasmic streaming). The recovery time of streaming (several minutes (35)) also outlasts the membrane potential pulse. If this view is correct, it will be possible to induce the deformation in absence of an AP. *(iii)* A third alternative is that *the cell surface deformation consists of two mechanical components* (one being the mechanical aspect of an AP and another triggered component). In the following, several potential mechanisms which could underlie the cellular deformation are discussed.

***Potential involvement of Ca²⁺ and actin-myosin motility in cell surface deformation.*** An increase in the intracellular Ca$^{2+}$ concentration has been implicated in many cellular events such as the stoppage of cytoplasmic streaming in Charophytes, muscle contraction, etc. It has been argued that the stoppage of cytoplasmic streaming during an AP in *Nitella* can be eliminated by various protocols that reduce the concentration of Ca$^{2+}$ in the extracellular medium (37, 38, 45). This would have allowed us to investigate the role of the events during streaming cessation for the cellular deformation. However, we and others (38, 45) found that these techniques cannot be readily applied to some *Chara* and *Nitella* cells. This apparent difference between Charophytes is surprising and – in itself – warrants further study. Nevertheless, there is some indication that muscle-like actin-myosin motility is not involved in the cell mechanical changes in *Chara*. The basis for this statement are experiments with cytochalasin D (Fig. 3), a substance that interferes with contractility in muscle cells (40). This substance also inhibits actin-myosin based cytoplasmic streaming in *Chara* (39). Under such conditions of impaired actin-myosin motility, however, the cell surface deformation persisted. This indicates that a different mechanism is at work. In any case, we suggest that future studies should investigate the cellular deformation during an AP in absence of other phenomena (cessation of cytoplasmic streaming, tonoplast AP, etc.).

***Cell surface deformation due to a change in transmembrane pressure***. Since hydrostatic pressure does not change during excitation of a cell, significant transmembrane pressure deviations should only arise if the chemical potential of water in the intra- or extracellular space is altered. The chemical potential depends on several parameters that define the thermodynamic state of water (temperature, concentration of solutes, etc.). To the best of our knowledge, non-ionic changes of the aqueous chemical potential during cell and tissue excitation have remained widely unexplored. It is often assumed that the solute concentration is the most likely parameter to change. Any variations of solute concentration in the intra- or extracellular compartments will lead to a chemical potential difference for water and an effective transmembrane pressure. This chemical potential difference will be equilibrated by transfer of water (*i.e.* osmosis) if the membrane is sufficiently permeable. In the classical theory of excitability, transmembrane flux of ions is a central premise. Therefore, cell surface displacements in neurons and plant cells have usually been interpreted as cell volume changes based on osmosis (25, 43). In nerve fibers, transmembrane flux of ions has been estimated from radiotracer studies to be on the order of $10^{-11}\ mol\ cm^{-2}\ impulse^{-1}$ (1, 7). In *Chara*, an influx of Ca$^{2+}$ and an efflux of $Cl^-$ and $K^+$ occurs during an action potential. The influx of Ca$^{2+}$ ($\sim 4 \cdot 10^{-14}\ mol\ cm^{-2}\ impulse^{-1}$) (36) is believed to be orders of magnitude smaller as compared to the efflux of $Cl^-$ and $K^+$ ($\sim 4 \cdot 10^{-9}\ mol\ cm^{-2}\ impulse^{-1}$) (46). The main consequence of these events in *Chara* should be an increase in extracellular osmolarity during an AP. This is equivalent to an increase in extracellular pressure ($p_{out}$). The cell could deform as a consequence of this transmembrane pressure difference. If the chemical potential difference for water is equilibrated by efflux from the cell, the pressure difference will decrease and in parallel the cellular volume will be reduced.

While a chemical potential difference for water could underlie the cellular deformations as observed herein, we deem it necessary to point out open problems. In the simplest case, *i.e.* if the cell membrane is highly permeable to water and if no significant transmembrane pressures act on the surface (43), one expects uniform shrinkage. Experimentally, however, inward as well as outward deflections of the cell surface were observed (Fig. S1). This indicates that a more comprehensive consideration of cellular geometry and of the mechanical forces that govern the surface is required. Another point relates to the aspiration experiments. The mechanical analysis indicated that an increase of $p_{out}$ can lead to the observed phenomenology (Fig. 4). However, if the efflux of solute is homogenous across the surface, $p_{out}$ will change inside as well as outside the pipette and thus the effective $\Delta p$ should remain unchanged. As a consequence, no new net force should act on the membrane. Another open problem regarding an ion-based osmotic mechanism of the deformation emerged from voltage clamp experiments by others (19). In voltage clamp it is assumed that transmembrane flux of ions can be monitored directly in the form of (ionic) currents (22). Ionic currents are low during hyperpolarization (~30 µA cm$^{-2}$) and high during depolarization (~1 mA cm$^{-2}$) (47). If ionic fluxes were the cause of the mechanical response in excitable cells, one would therefore expect larger displacements during depolarization as compared to hyperpolarization. However, the opposite was observed (19). This constitutes a central contradiction that will have to be addressed.

Finally, it must be emphasized that an osmotic mechanism implies that in most cases (*i.e.* when the hydraulic permeability of the membrane is nonzero) a change in volume will take place. To the best of our knowledge, however, there does not exist sufficient evidence to claim that cell volume changes during an AP. Up to date, one study with Charophytes has indicated that volume is slightly decreased (32). However, more experimental evidence, in particular from neuronal preparations, is required. In the meanwhile, it is expedient to also consider mechanisms of deformation at widely constant cell volume. As an analogy, one may imagine a droplet of water sitting on a surface. The droplet will deform at constant volume if, for instance, its surface tension is changed by addition of a small amount of surfactant.

***Cell surface displacements due to changes in surface tension and/or bending rigidity.*** In a 1945 paper, Hodgkin and Huxley considered several potential mechanisms of the nerve action potential (48). One implied a cooperative change of orientation of lipid dipoles. This possibility was rejected, because such a process should notably alter the electrical membrane capacitance ($C_m$) – a parameter that at that time was assumed to be constant during an AP. In subsequent studies, it was demonstrated, however, that the assumption of constancy of $C_m$ had been premature (49). This may have led to misinterpretations, for example, because dynamic capacitive currents were ruled out in the original works (11). Based on this point and several others, a criticism of the electrical theory was formulated by Kaufmann (11, 12) and more recently by Heimburg (13, 14). These authors proposed that an AP is a pulse propagating in the quasi 2-dimensional membrane interface. Such a reversible (adiabatic) phenomenon must be associated with transient changes in forces and fluctuations of all thermodynamic observables of the system (electric field, pressure, temperature, surface area etc.). More recently, others have elaborated on this proposition (13, 14, 50–52). It was demonstrated that linear (50) as well as non-linear, self-stabilizing pulses (solitary waves) (51, 52) can be excited in lipid monolayers – the simplest model system of a cell membrane. These pulses indeed manifest in *all* thermodynamic variables (*e.g.* electrical, thermal, optical, etc.) (53–55). This is also the case for APs (1, 2, 10, 15, 19). If one compares the solitary waves in protein-free lipid monolayers at the water-air interface with action potentials, additional similarities exist (threshold, amplitude saturation, etc.). Thus, the thermodynamic theory predicts a lateral pressure pulse (related to the surface tension) as well as a change in mechanical susceptibilities (area compressibility and bending rigidity) during cellular excitation. The biphasic mechanical changes during an AP in *Chara* (Figs. 2—4) may be the consequence of pulse-associated changes in $\sigma$ and/or $\kappa$ of the excitable medium. The present analysis of micropipette aspiration results suggests that the observed phenomenology (Fig. 4) could be explained by a sufficient decrease in surface tension (by ~50 %) and/or bending rigidity (by ~2 orders). Such changes are large but not unrealistic, as a decrease in surface tension by ~10% was demonstrated in lipid monolayer pulses (50) and during phase transitions in lipid bilayers $\kappa$ can be reduced by 1—2 orders of magnitude (56). For most state changes, however, $\sigma$ and $\kappa$ will change *simultaneously*. For example, when a fluid lipid membrane is compressed isothermally into the phase transition regime, the lateral pressure as well as $\kappa$ increase. In order to understand the particular relations between $\sigma$ and $\kappa$ in *Chara*, it will be necessary to obtain state diagrams of the excitable membrane. It will also be important to understand the sequence of state changes that the excitable membrane undergoes during an action potential. Finally, the permeability of a membrane depends on its thermodynamic state and thus any state change, whether induced isothermally or adiabatically, will lead to the realization of a different permeability (57). Thus, a propagating state change in the plasma membrane (*i.e.* a density pulse) could entail transmembrane transport phenomena.

## Conclusion

The present work demonstrated that an AP is associated with a significant surface deformation (~1—10 μm) in plasmolysed *Chara* cells. This deformation co-propagates with the electrical signal, is biphasic (72% of cases) and reversible. Due to the magnitude of the displacements as well as the slow time scales of the pulse, this preparation is well suited to study the physical origin of cell mechanical changes during excitation. Falsifiable predictions were made concerning the surface property changes ($p_{out}, \sigma, \kappa$) that may be involved. Future work should aim at understanding the coupling between the electrical and the mechanical signal in Charophytes and neurons. Finally, it will be important to investigate if interfacial pulses in membrane models (lipid monolayer) and action potentials in cells are based on the same physical principles.

## Authors contributions

## Acknowledgements

We wish to thank K. Kaufmann for talks and personal discussions, which have motivated us to conduct the present experiments. The interested reader is particularly encouraged to consult K. Kaufmann's writings from the 1980s. We also acknowledge discussions with S. Shrivastava and crafting of measurement chambers by D. Campbell. CF is grateful for funding by the Max Kade Foundation (http://maxkadefoundation.org/). MFS would like to acknowledge financial support by the German Science Foundation (DFG) as well as the research unit SHENC.